# Data Acquisition Software for CBM-TOF super module quality control

Jiawen Li, Xiru Huang, Ping Cao, Chao Li, Jianhui Yuan, Wei Jiang, Junru Wang, Qi An

*Abstract*—The time-of-flight (TOF) system in the Compressed Baryonic Matter (CBM) experiment is composed of super modules based on multi-gap Resistive Plate Chambers (MRPC) for high-denseness, high-resolution time measurement. In order to evaluate the quality of detectors during the mass production, a distributed data readout system is developed to meet with the high data rate of 6.4Gbps, and a related data acquisition (DAQ) software is implemented under Linux operating system to test and verify the feasibility of the distributed data readout method. In this paper, the DAQ software is focused on data collection, event building, status monitoring and system controlling. Laboratory tests confirmed the function of the DAQ software and show that the overall data transfer rate of a single data transmission path can reach up to about 550Mbps which already meet the demand.

## I. Introduction

CBM experiment at the Facility for Antiproton and Ion Research (FAIR) aims at the search for phase transitions in the phase diagram of strongly interacting matter as well as the study of strange and charmed particles production [1][2]. Charged hadron is identified in CBM by a TOF super module system placed 10m downstream of the fixed target [3]. For the purpose of quality control of CBM-TOF super module, a 320-channel time digitizing and readout electronic system shown in Fig.1 is designed. To read out data in real time, the system has a distributed architecture, including front-end electronics (FEE), back-end electronics (BEE) and DAQ software.

FEE mainly contains 10 TDC for time digitizing and 1 TDC Readout Motherboard (TRM) for reading out multiple TDC. BEE is mainly composed of 16 Data Readout Modules (DRM) for data forwarding and 1specific DRM for status and command routing in the PXI crate, as well as a clock & trigger system (CTS) for clock and trigger distribution. Each DRM is

Manuscript received June 23, 2018.
This work was supported by the National Basic Research Program (973 Program) of China under Grant 2015CB856906.

The authors are with the State Key Laboratory of Particle Detection and Electronics, University of Science and Technology of China, Hefei 230026, China (email: xiru@ustc.edu.cn).
J. Li, P. Cao, J. Yuan and W. Jiang are with the Department of Engineering and Applied Physics, University of Science and Technology of China, Hefei 230026, China.
X. Huang, C. Li, J. Wang and Q. An are with the Department of Modern Physics, University of Science and Technology of China, Hefei 230026, China.

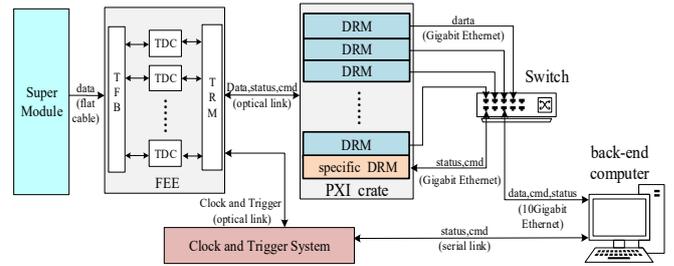

Fig. 1. Structure of the distributed data readout system

based on the system on chip (SOC) and the Ethernet techniques, so that data can be parallel transmitted to the back-end computer. The total data rate of the readout electronics system with 16 DRM is about 6.4Gbps.

Accordingly, a Data Acquisition software is basically designed to transmit data in real time. The software also has a Graphical User Interface and can provide functions of event building, command distribution and online status monitoring.

## II. Software Architecture

The functions required for the data acquisition system in the particle physics experiment generally include two aspects: one is to transmit data obtained from the FEE to the back-end computer; the second is to send command and configuration to the FEE and BEE, and feed back the operating status of the FEE and BEE to the back-end computer. The Data Acquisition Software that focuses on the function of the first aspect has a distributed and hierarchical architecture, and consists of three parts connected with Ethernet as shown in Fig.2: Data Forwarding Node (DFN), Data Aggregation Node (DAN) and Graphical User Interface (GUI).

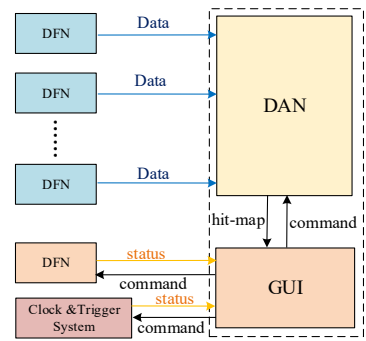

Fig. 2. Architecture of DAQ software system

Data Forwarding Node aims at forwarding data to Data Aggregation Node or transmitting status and commands with GUI. DAN and GUI both run on the back-end computer. DAN is mainly responsible for data receiving and event building. GUI provides friendly and interactive interface for users to control and monitor the electronics system. DAN and GUI also contains online data analysis and hit-map display to evaluate the quality of detector. Such layered design is easy to upgrade, as the number of DFN can be configured according to the requirement of experiment, which is suitable for the distributed readout system, and GUI can be customized without any code modification. In this way, the data readout channel is separated from the system status and the control command transmission channel, not interfering with each other.

### III. IMPLEMENTATION OF DFN

There are two types of DFN: the first is a distributed data readout node responsible for data transmission which runs on DRM, and the second is responsible for parameter configuration, command control, and status monitoring, which runs on the specific DRM. The structure of DFN is shown in Fig.3. DFN only forwards data, status and command it receives instead of processing them.

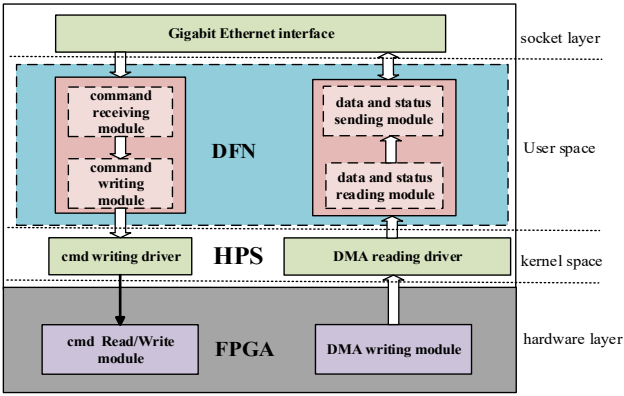

Fig. 3. Structure diagram of DFN

DRM based on Cyclone V SoC is composed of two distinct portions: a hard processor system (HPS) and a Field-Programmable Gate Array (FPGA), which uses the system-on-a-chip technology to realize the Gigabit network transmission capability of the readout subnode. The FEE data interface is a logical interface that is uniformly defined. The interface module of FEE caches data and sends it to a data packaging module for processing. The packaged data is sent to the DAQ interface module, which uses TCP/IP to send and receive data. The client-server model is usually used for network programming. Here, DFN and GUI acts as client and DAN acts as server.

Data Forwarding Node is designed as a concurrent software based on multithreading technology. Using multithreading techniques, the program is divided into multiple independent tasks that increase the response speed [4]. There are three threads in DFN: main thread, data transceiving thread, command transfer thread. Main thread connects client socket to server and creates other threads with detached attributes. Command transfer thread send the command for configuring and hardware controlling to FPGA, using the write method defined by the driver of HPS-to-FPGA interface. The task of the data transceiving thread is to receive data from FPGA via the Direct Memory Access (DMA) driver of the FPGA-to-HPS interface and then transmit them to PC via the Ethernet.

Tests show that the FPGA-to-HPS interface is the bottleneck of transmission rate, so the DMA transfer efficiency is critical. Therefore, the DMA driver and corresponding FPGA logic module is optimized to make DMA transfer and kernel interrupt response executed in parallel. Meanwhile, the data transceiving thread in DFN application uses the mmap method to reduce time of copying data from kernel space to user space and improve transmission efficiency.

### IV. IMPLEMENTATION OF DAN

DAN runs on the back-end PC and transmits various data with DFN and GUI through network. In DAQ software, DAN serves as server. The Structure diagram of DAN is shown in Fig.4.

Data Aggregation Node also utilizes multithreading technology, comprising four parts: main thread, data collection thread, data saving thread, command processing thread, online analysis thread. Main thread initializes the mutex, the list of threads and the list of event building buffer, creates TCP sockets and accepts client connection requests from DFN and GUI, then create data collection threads and data saving thread. Data collection threads, the number of which depends on the number of DFN, receives data from DFN, and writes data to the list of event building buffer according to the trigger id number. List data structure shown in Fig.5 is utilized for event building. The number of nodes in the list depends on the data rate and the processing capacity of the data saving thread. All data collection threads cooperate with each other to complete event building in the list, and notify data saving thread to read the assembled data via condition variables. There's only one data saving thread which will be wakened up by the condition variable, then save assembled data to file. From the perspective of the producer-consumer model, data receiving thread is the producer and the data saving thread is the consumer. Online analysis thread read assemble data saved in the file and count the hit in each position, then send the hit-map to GUI.

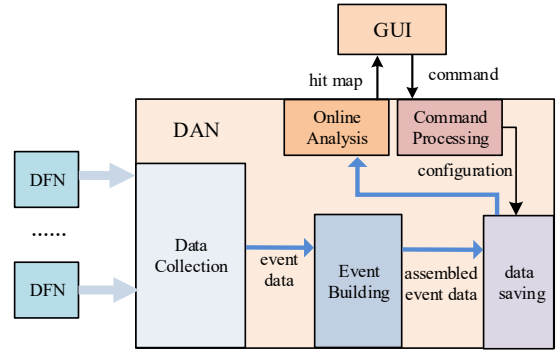

Fig. 4. Block diagram of DAN

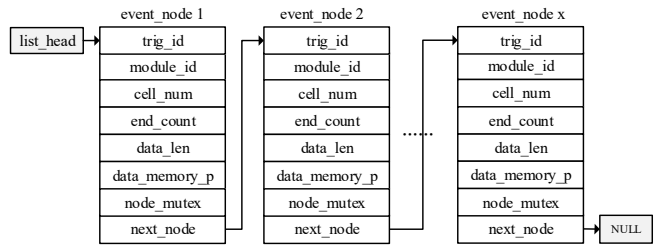

Fig. 5. List structure and event building process

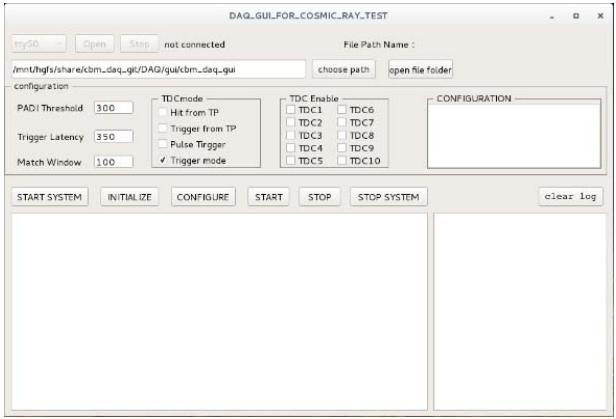

Fig.6. GUI for control and monitor.

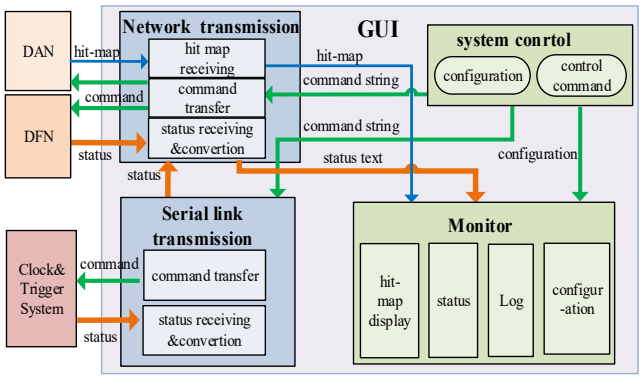

Fig.7. Structure diagram of GUI

## V. GUI

GUI shown in Fig.6 is programmed with Qt language [5], also uses multithreading technology to improve the response rate of the interface application. It consists of GUI main thread, transfer thread and hit-map display thread. GUI main thread displays status and interacts with users. When user pushes command buttons, GUI main thread creates corresponding strings and send them to transfer thread. Transfer thread receives status data and convert it to text message, then send to GUI main thread. In addition, it responds to GUI main thread command signal, sending command string to DFN, DAN and CTS. Hit-map display thread reads the hit-map information from DAN, then display it in the histogram so that the performance of the detectors can be directly reflected [6]. The Structure diagram of GUI is shown in Fig.7.

## VI. TRANSMISSION TEST

As mentioned before, the total data rate of the readout electronics system is about 6.4 Gbps, and the network transmission rate is supposed to reach up to 400 Mbps for a single data transmission path. Therefore, the network transmission performance of the system is particularly important.

Tests have been done in laboratory to verify feasibility and performance of the DAQ software, where FPGA in TRM generates incremental code as data source. The test software basically uses the architecture of DAQ software in this article, except that the test version of DAN adds rate calculation section to data collection module. Timestamps are made in each transmission, and the amount of data transmitted is accumulate at the same time, so that the transmission rate could be calculated once every certain time. Besides, Bit Rate Error (BER) verification is added, in which the data saved in the file are verified as incremental code.

There's no any error occurred during the test last for one hour. Fig.8. shows the result of transmission rate. The average rate for a single data transmission path is 550Mbps and the standard deviation is 29Mbps, which fully satisfies the readout system requirements.

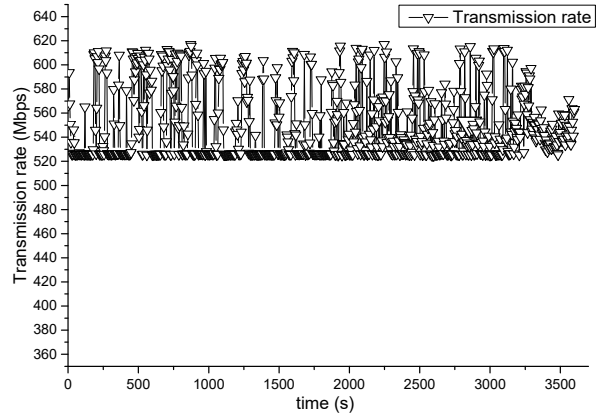

Fig.8. Transmission rate for a single data transmission path

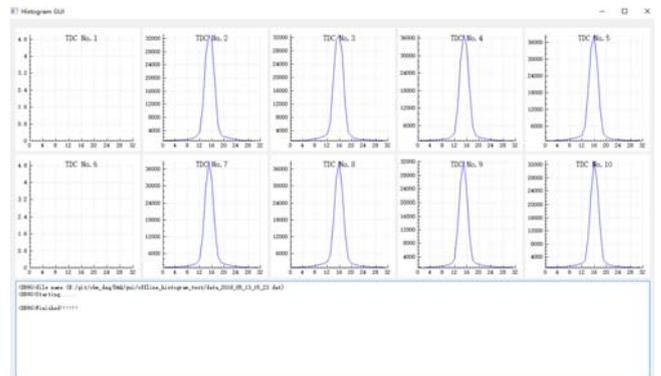

Fig.9. Hit-map display in cosmic ray test

Cosmic ray test with MRPC detectors is also carried out. In the first stage, the signals are digitized by eight TDC boards.

The online analysis in DAQ software worked well as the hit-map shown in Fig.9.

## VII. Conclusion

The Data Acquisition Software for quality control of CBM-TOF super module detector can run stably, and it has a hierarchical structure that is easy to extend. Meanwhile, it meets the readout demand well as laboratory test shows that a single data transmission path achieved approximate 550Mbps data transfer rate in the case of full link from FEE to backend computer storage.

## References

[1] J. M. Heuser, "Status of the CBM experiment," in EPJ Web of Conferences, 2015, vol. 95, p. 01006: EDP Sciences.
[2] P.-A. Loizeau, "Development and test of a free-streaming readout chain for the CBM Time of Flight Wall," 2014.
[3] I. Deppner et al., "The CBM time-of-flight wall," Nuclear Instruments and Methods in Physics Research Section A: Accelerators, Spectrometers, Detectors and Associated Equipment, vol. 661, pp. S121-S124, 2012.
[4] N. Matthew and R. Stones, "POSIX Threads," in *Beginning Linux Programming*, 4th ed. Indianapolis, Indiana: Wiley, 2008, ch.12, pp.495-524.
[5] J. Blanchette and M. Summerfield, *C++ GUI Programming with Qt 4*, 2nd ed. Prentice Hall, 2008.
[6] F. Li, K. Zhu, J. Zhao, L. Wang, and Y. Liu, "Design and implement of BESIII online histogramming software," *Nuclear Electronics & Detection Technology*, vol.27, no.3, pp.462-465, May. 2007.